\documentclass[12pt,psfig,epsfig,graphicx,psfrag]{article}
\usepackage{graphicx}
\usepackage{amsfonts,amssymb,amsmath}
\usepackage{hyperref}

\newcommand{\be}{\begin{equation}}
\newcommand{\ee}{\end{equation}}
\newcommand{\ba}{\begin{eqnarray}}
\newcommand{\ea}{\end{eqnarray}}

\def\L5{\tilde{\Lambda}}

\newcommand{\I}{\mathrm i}
\newcommand{\m}{\hat m}
\def\pd{\partial}
\def\a{\alpha}
\def\b{\beta}
\def\g{\gamma}
\def\di{\mathrm{d}}

\def\m{\mu}
\def\n{\nu}

\def\l{\lambda}

\def\r{\rho}
\def\s{\sigma}
\def\e{\epsilon}
\def\d{\delta}
\def\WTDiff{\rm WTDiff}
\def\WRS{\rm WRS}
\def\RS{\rm RS}
\def\pdi{/\hspace{-.2cm}\pd}

\renewcommand{\di}{{\mathrm{d}}}

\def\1{\mathchoice{\rm 1\mskip-4.2mu l}{\rm 1\mskip-4.2mu l}%
{\rm 1\mskip-4.6mu l}{\rm 1\mskip-5.2mu l}}

\begin{document}
\begin{flushright}
\end{flushright}
\vskip 1cm

\begin{center}
{\Large {\bf Transverse Symmetry and Spin-$3/2$ Fields}}\\[1cm]
D.~Blas$^a$\footnote{dblas@ffn.ub.es},\\ 
$^a${\it Departament de F\'isica Fonamental \ \&
\ Institut de Ci\`encies del Cosmos,\\
 Universitat de Barcelona,
Diagonal 647, 08028 Barcelona, Spain.}\\
\end{center}
\vskip 0.2cm

\noindent
\begin{abstract}
We study the possible covariant Lagrangians that describe the propagation
of pure spin-$3/2$ particles. We show that, apart from the well-known Rarita-Schwinger Lagrangian,
there is another possibility where the field is described by a $\g$-traceless combination
and that both Lagrangians yield the same physical predictions for the interaction of
 conserved sources.
We also prove that for the case when the spin-2
field is described by a traceless field, there is no possible spin-$3/2$ action
that makes the system supersymmetric. Nevertheless,
the interaction between this field and the spin-$3/2$ field may be possible.
\end{abstract}



\newpage

\section{Introduction}

It has been recently shown that the free massless spin-2 field
can be consistently described by a Lagrangian depending on the
traceless tensor field
\be
\hat h_{\m\n}=h_{\m\n}-\frac{1}{4}\eta_{\m\n}h,
\ee
and endowed with a
{\em reducible} gauge symmetry \cite{Alvarez:2006uu,Blas:2007pp}. This
formulation is in some sense opposite to the standard approach of
higher spin, which resorts to the introduction of auxiliary fields to
build a covariant Lagrangian which yields the correct equations of
motion \cite{Fierz:1939ix,Fronsdal:1978rb,Fang:1978wz,deWit:1979pe}
(see also \cite{Singh:1974qz,Singh:1974rc} for the massive case).
The analysis has been extended to bosonic fields of higher spin in
\cite{Skvortsov:2007kz}. In general, even if the Lagrangians for the
traceless field are related to the original Fronsdal model of
\cite{Fronsdal:1978rb}, the equivalence between both formulations is
not trivial. In fact,  a naive counting
 indicates that the Lagrangian of the traceless field has an
extra propagating degree of freedom. However, as shown in
\cite{Skvortsov:2007kz}, the appearance of a {\em tertiary}
constraint in this description kills the extra
degree of freedom and makes
both theories equivalent at the classical level (except
for an integration constant).
This phenomenon is reminiscent of the reason why {\em
unimodular} gravity is equivalent to General Relativity  \cite{Henneaux:1989zc}.

The covariant
description of {\em fermionic} fields of spin
$s>1/2$ also needs the introduction of auxiliary fields which are
rendered spurious by an associated gauge symmetry
\cite{Fang:1978wz}. A natural question one may ask is whether, as
happens in the bosonic case, there exists more than one Lorentz invariant
 Lagrangian that describes the propagation of the higher spin particles.
 Besides, from the analogy with the bosonic case,
 one may expect that the new Lagrangian depends
on a field with fewer degrees of freedom than in the standard
formulation.
One of the  purposes of this work is to study this possibility
for the spin-$3/2$ field.

Besides, it is well known that the standard spin-$2$ and
spin-$3/2$ actions constitute  a supersymmetric action
\cite{VanNieuwenhuizen:1981ae}.
 This property is crucial in finding an
action which couples the spin-$3/2$ field {\em consistently}\footnote{
By consistent coupling we mean that the coupled theory is invariant
under a deformation of the gauge symmetry of the free theory. This is
understood as a necessary condition for consistency as the gauge symmetry
is essential to make the theory {\em unitary}.} to the
graviton and we will look for
a similar minimal
supersymmetric extension for the traceless case.\\

 The paper is organized
as follows. In Section \ref{RSsection},
 we will present the most general Lorentz invariant
 Lagrangian for
 the Majorana spin-$3/2$ in terms of the spinor-vector $\psi_\m$,
and identify the possible gauge symmetries of the
action. In Section \ref{sectionprop}, we will study the propagators
of the Lagrangians which are endowed with a gauge symmetry and
comment on the possibility of consistently coupling the
field to the electromagnetic field. In Section \ref{sectionSugra},
we will study the possible supersymmetric linear actions
for the traceless spin-$2$ field. We will finally present
our conclusions and outlook.

We will follow the conventions of \cite{deWit:1985aq}\footnote{We will work with
\be \eta_{\m\n}=\{+,-,-,-\},\quad \{\g^\m,\g^\n\}=2\eta^{\m\n},
\quad \g_5=\I\g^0\g^1\g^2\g^3, \quad \e^{0123}=1. \ee}  and work
 with a Majorana
vector-spinor $\psi_\m$.

\section{Lagrangians for Pure Massless Spin-$3/2$}\label{RSsection}

The most general local, Lorentz invariant Lagrangian for a massless Majorana
 vector-spinor $\psi_\m$ and involving just
 first order derivatives is given by\footnote{For
  a Dirac spinor, the coefficients in front of the first and
second terms do not necessarily coincide.}
\be \label{gaction}
{\mathcal S}^{(3/2)}=\int \di^4 x \ \bar \psi_\m \left(\lambda
(\g^\m \pd^\n + \g^\n \pd^\m)+\vartheta \g^\m \pdi \g^\n+\zeta
\eta^{\m\n}\pdi\right)\psi_\n.
\ee
After a transformation of the
form
\be \label{conftr} \psi_\m\mapsto \psi_\m- \frac{a}{4}
\g_\m\g^\r\psi_\r,
\ee
the coefficients change as
\be
\label{coefftr} \l\mapsto \l\left(1-a\right)-\frac{a}{2}\zeta, \quad
\vartheta\mapsto
\vartheta(1-a)^2-\frac{a(1-a)}{2}\l+\frac{a}{2}\left(1-\frac{a}{4}
\right)\zeta.
\ee
This transformation is a field redefinition which
makes one of the coefficients spurious  except for the case $a=1$.
In this pathological case, the transformation is {\em not}
invertible.

The Majorana field $\psi_\m$ has 16 real independent components, all
of which will be dynamical for a general action of the form
(\ref{gaction}). However, if the action is to describe {\em just} a massless
particle of spin-$3/2$, only the $\pm3/2$ polarizations should be dynamical, which
implies the need for a gauge symmetry
to render the remaining polarizations non-dynamical\footnote{Recall also that
fermions have half as many propagating degrees of freedom as components, since the other half are
canonical momenta.}.
In fact, the
standard action for spin-$3/2$, known as  Rarita-Schwinger (RS) and characterized
by the values
$\l=-\vartheta=-\zeta$ \cite{Rarita:1941mf} (and the coefficients related to it by a
transformation (\ref{coefftr}) for $a\neq 1$),  is invariant under the
transformation
\be
 \psi_\m\mapsto \psi_\m+\pd_\m \e,
\ee
where $\e$ is a Majorana spinor.
This
transformation can be generalized to
\be \label{generalgauge3/2}
\psi_\m\mapsto \psi_\m+\pd_\m \e+\g_\m \varphi,
\ee
which is the
most general covariant gauge transformation for the field $\psi_\m$ which
does not involve the spin-$3/2$ components of the field.  Under
these transformations, the general
 action  (\ref{gaction}) changes as
\ba
\d S^{(3/2)}=-2\int \di^4 x \Big(\{(\l+\vartheta)\Box
\bar\e&+&(\l+4\vartheta-\zeta)
\pd^\a \bar \varphi \g_\a\}\g^\m \psi_\m\nonumber\\
&& -\{(\l+ \zeta)\pd^\a \bar\e\g_\a+2(2
\l+\zeta)\bar\varphi\}\pd^\m\psi_\m\Big).
\ea
For $2\l+\zeta\neq0$ the previous variation cancels for
\be
\bar
\varphi=-\frac{(\l+\zeta)\pd^\a\bar\e\g_\a}{2(2\l+\zeta)},\quad
(3\l^2+2\zeta\l+\zeta^2- 2\vartheta\zeta)\Box\bar \e=0.
\ee
In other words, for
\be
\label{RSrelated}
\vartheta=\frac{\zeta^2+2\zeta\l+3\l^2}{2\zeta}, \quad 2\l+\zeta\neq0,
\ee
the action (\ref{gaction}) is invariant under (\ref{generalgauge3/2})
with
$$\bar\varphi=-\frac{(\l+\zeta)\pd^\a\bar\e\g_\a}{2(2\l+\zeta)},$$ and
$\e$ remains a {\em free} Majorana spinor.
 As it is clear from
(\ref{coefftr}), not all these possibilities are independent. In fact,
they are {\em all} equivalent to the  RS action
after a field redefinition with parameter
$$a=\frac{2(\l+\zeta)}{\zeta}.$$
For the singular case $2\l+\zeta=0$, the
variation of the action cancels provided that
\be \pdi \e=0, \quad
(\l+4\vartheta-\zeta)\pd^\a\bar\varphi\g_a=0.
\ee
In this case,
the Majorana spinor $\varphi$ will be free if
$ \l=\zeta-4\vartheta$,
which, together with $2\l+\zeta=0$,
implies that
\be
\l=-\frac{1}{2}\zeta, \quad \vartheta=\frac{3}{8}\zeta.
\ee
Substituting the previous values in (\ref{gaction})
(and fixing $\zeta$), one finds the
action
\be
\label{WRSaction} {\mathcal S}^{(3/2)}_{\WRS}\equiv{\mathcal
S}^{(3/2)}_{\RS}(\hat \psi_\m)= -\frac{1}{2}\int \di^4 x\ \bar{\hat
\psi}_\m\e^{\m\n\r\s}\g_5 \g_\n \pd_\r \hat \psi_\s,
\ee
where ${\mathcal
S}^{(3/2)}_{\RS}$ refers to the RS action and $\hat
\psi_\m\equiv\psi_\m-\frac{1}{4}\g_\m\g^\a\psi_\a$. Thus, this
 action corresponds to the singular transformation
(\ref{conftr}) with $a=1$. This means that
the action is written in terms of a covariant
quantity with {\em fewer} degrees of freedom than the original
field $\psi_\m$. Indeed, $\hat \psi_\m$ is $\g$-traceless,
$$
\g^\m\hat\psi_\m=0,
$$
which means that $\hat\psi_\m$ has 12 independent real components.
The WRS label comes from
the analogy of the transformation (\ref{generalgauge3/2})
involving the field $\varphi$ (known as special supersymmetry, or
simply, $S$-symmetry \cite{Fradkin:1985am}) with the Weyl gauge
symmetry. Besides, the first
condition in (\ref{RSrelated}) is also satisfied in this case.

As happens in the spin-2 case,
for generic $\e$ and $\varphi$ there is no action in the family
(\ref{gaction}) invariant under the most general possible gauge
symmetry\footnote{Similarly to what happens to the Weyl symmetry for
theories invariant under diffeomorphisms,
 an action invariant under this gauge group exists once
higher derivatives terms are included,
but the theory is not unitary (see \cite{Fradkin:1985am}).}. In particular, this means
that some of the low spin components of the field $\psi_\m$
may be dynamical as they are not automatically killed by the gauge symmetry.
To our knowledge, the action (\ref{WRSaction}) has not been studied in detail
in the past (for the RS action see, {\em e.g.},
\cite{VanNieuwenhuizen:1981ae,deWit:1985aq}).

In terms of $\hat\psi_\m$, the
action is invariant under the transformation
\be
\delta \hat\psi_\m=\pd_\m\e,
\ee
satisfying $\pdi \e=0$.
If we, instead, consider  $\psi_\m$ as the fundamental
field and think of $\hat\psi_\m$ as a derived quantity, the
action is invariant under (\ref{generalgauge3/2}) with $\pdi \e=0$.
It is important to note that the action (\ref{WRSaction}) is {\em not}
related to the RS action by a
gauge fixing term, as the only covariant gauge fixing term just
involves the term proportional to
$\vartheta$  in (\ref{gaction}). \\

There is another way in which we can show that the RS and the WRS actions
are the only possibilities out of the general action (\ref{gaction}) endowed
with a gauge symmetry. Namely, we can analyze the  structure
of the equations of motion of the
theory and find all the degenerate possibilities. To this aim, it is
convenient to decompose the field $\psi_\m$ into irreducible
representations of the $SO(3)$ group (see {\em e.g.} \cite{Deser:1977ur}),
\be
\psi_0=A, \quad \psi_i=t_i+\g_i \chi+\pd_i E,
\ee
with $\g_i
t^i=\pd_i t^i=0$. The presence of the $\g_i$ matrices in
the definition of $\chi$ implies that it is an anti-Majorana
fermion\footnote{We can also define $\chi=\g_0\eta$ with $\eta$
being a Majorana spinor.}
$$
\bar\chi=-\chi^TC.
$$
In terms of the previous fields, the general Lagrangian (\ref{gaction})
 can be decomposed into a spin-$3/2$ and a spin-$1/2$ part,
\be
\label{decompoaction}
{\mathcal L}={\mathcal
L}^{(3/2)}+{\mathcal L}^{(1/2)},
\ee
with ${\mathcal L}^{(3/2)}\equiv-\zeta\bar t_i \pdi\ t_i$ and
\ba {\mathcal
L}^{(1/2)}&&\equiv\bar E\left\{(\zeta-\vartheta)\g_0\pd_0-
(2\l+\vartheta+\zeta)\g_i\pd_i\right\}\Delta E+\bar A\left\{
(2\l+\vartheta+\zeta)\g_0\pd_0-\g_i\pd_i(\zeta-\vartheta)\right\}A
\nonumber\\
&&+\bar
\chi\left\{3(3\vartheta-\zeta)\g_0\pd_0-\g_i\pd_i(6\l+9\vartheta-
\zeta)\right\}\chi+2\bar\chi\left\{-(4\l+3\vartheta+\zeta) \Delta
E-(3\vartheta-\zeta)\g_0\g_i
\pd_0\pd_i E\right\}\nonumber\\
&&+ 2\bar A\left\{-(\l+3\vartheta)\g_0\g_i\pd_i
\chi+(\l+\vartheta) [\pd_0(3\chi-\g_i\pd_i E)-\g_0\Delta
E]\right\}.\nonumber
\ea
The presence of a gauge symmetry can be identified by the
singular character of the kinetic term of the
equations of motion \cite{Henneaux:1992ig}.
The kinetic term can be written as
\begin{displaymath}
\left(\begin{array}{lll}
(\zeta-\vartheta)\g_0\Delta & \ (\zeta-3\vartheta)\g_0\g_i\pd_i&\ (\l+\vartheta)\g_i\pd_i\\
(\zeta-3\vartheta)\g_0\g_i\pd_i&\ 3(3\vartheta-\zeta)\g_0&\ 3(\l+\vartheta)\\
-(\l+\vartheta)\g_i\pd_i&\ 3(\l+\vartheta)&\
(2\l+\vartheta+\zeta)\g_0
\end{array}\right)
\left(\begin{array}{c} \dot E\\ \dot \chi \\ \dot A
\end{array}\right).
\end{displaymath}
The determinant of the matrix multiplying the
time derivative of the fields is
\be 16\zeta^4(-2\vartheta
\zeta+\zeta^2+2\zeta\l+3\l^2)^4\Delta^4.
\ee
Thus, we find that the
theory will include constraints whenever (we take $\zeta\neq0$ as
otherwise the spin-$3/2$ degrees of freedom are not present)
\be
\vartheta=\frac{\zeta^2+2\zeta\l+3\l^2}{2\zeta}.
\ee
As we found
previously, this condition corresponds to the existence of a gauge
symmetry of the form (\ref{generalgauge3/2}). The previous
decomposition can be used to identify the Lagrange multipliers
and constraints of the theory. Once the constraints are
introduced back into the action (\ref{decompoaction}),
the only remaining fields are the propagating degrees of freedom.
\\

\section{Propagator and Coupling of  the WRS Lagrangian}\label{sectionprop}

The analysis of the general Lagrangian (\ref{gaction}) is beyond the scope of this work.
The absence of a gauge symmetry means that all the degrees of freedom propagate in general
and one expects the presence of ghosts in the low-spin states \cite{VanNieuwenhuizen:1981ae}. For the
RS Lagrangian, the propagator, spin content and unitarity properties can be found in
\cite{Sterman:1977ds,Das:1976ct,VanNieuwenhuizen:1981ae}.
In this case, the fact of having a gauge
symmetry involving a derivative allows to kill all the low-spin states, leaving
just the $\pm 3/2$ components \cite{Sterman:1977ds}.

 For the WRS action
the naive counting of propagating degrees of freedom implies the presence of  spin-$1/2$ components.
Let us show that this is indeed the case.  From the action (\ref{WRSaction}),
one readily finds that the equations of
motion for $\psi_\m$ are the $\g$-traceless part of the RS case in
the gauge $\g^\m\psi_\m=0$. This gauge can be easily reached in both
the WRS and
RS case. One finds,
\be
{\mathcal R}_{\WRS}^\m\equiv\frac{\d {\mathcal
L}_{\WRS}}{\d \bar\psi_\m}=
\left(\d^\m_\a-\frac{1}{4}\g^\m\g_\a\right) \frac{\d {\mathcal
L}_{\RS}(\hat\psi_\m)}{\d \bar{\hat \psi}_\m}\equiv
\left(\d^\m_\a-\frac{1}{4}\g^\m\g_\a\right){\mathcal
R}_{\RS}^\a(\hat\psi_\m)=0,
\ee
with $\g_\a {\mathcal R}^\a_{\WRS}=0$,
which is the Bianchi identity associated to the
 $S$-symmetry. Contracting the equations of
motion with the derivative operator,
one finds
\be
\pd_\m {\mathcal
R}^\m_{\WRS}=-\frac{1}{4}\pdi\left(\g_\a{\mathcal
R}^\a_{\RS}(\hat\psi_\m)\right).
\ee
Thus, contrary to
what happens in the bosonic case (cf.\cite{Alvarez:2006uu}),
we do not recover the missing
equations of the RS Lagrangian (in this case the $\g$-trace of the RS
equations of motion). This
result was expected as in the case under study there is no gauge
invariance left in the WRS action for the $\hat\psi_\m$ field,
 which means that no secondary or tertiary
constraints can appear. In fact, from the identity
$$
\g_\a{\mathcal R}^\a_{\RS}(\hat\psi_\m)=-2\pd^\a\hat\psi_\a,
$$
we see that there is a spin-$1/2$ propagating degree of freedom as the equation of motion for
$\pd^\a\hat\psi_\a$ is
\be \label{newspin} \pdi\pd^\a\hat\psi_\a=0,
\ee
in contrast to the RS case where $\pd^\a\hat\psi_\a$ cancels on
shell\footnote{ These equation of motion are also obtained if we add
a gauge fixing term
\be \l \bar \psi_\m\g^\m \g^\n \psi_\n
\ee
to the RS action.
This is reminiscent to what happens in unimodular gravity
\cite{Henneaux:1989zc}.}.
Besides,  the residual gauge
transformation leaves this combination invariant as
\be \d
\pd^\a\hat\psi_\a=\Box \e=0.
\ee
The previous arguments imply that the free WRS case is
not in general equivalent to the RS case as there is one more
spin-1/2 propagating degree of freedom. However, notice that
if we fix the initial condition
\be
\label{boundaryc}
\pd^\a\hat
\psi_\a|_0=0,
\ee the equation (\ref{newspin}) implies that the missing
equations also hold and that both systems are equivalent. This
situation is analogous to what happens in gauge invariant theories when
one fixes the gauge through a covariant quadratic gauge fixing
term \cite{Das:1976ct,Itzykson:1980rh}.
 Thus, we conclude that both actions have the same equations
 of motion
  once we impose the (\ref{boundaryc}) initial condition in the WRS case.


To show that both theories yield equivalent physical results,
we can couple the free spin-$3/2$ field to a conserved
source $J_\a$, with $\pd^\a J_\a=0$, and find the propagator
that mediates the interaction between two sources.  The
most general covariant non-derivative coupling will be of the form
$$
{\mathcal S}_{int}=\int \di^4x
\bar\psi_\m\left(J^\m-\frac{b}{4}\g^\m\g_\a J^\a\right)+h.c.
$$
The consistency of the equations of motion implies that for the RS
case $b=0$, whereas for WRS $b=1$. The equations of motion for the
WRS case are
\be \label{eom}
\left(\d^\m_\a-\frac{1}{4}\g^\m\g_\a\right)\left({\mathcal
R}_{\RS}^\a(\hat\psi_\m) -J^\a\right)=0.
\ee
Again, from the
conservation of the current and the Bianchi identity for ${\mathcal
R}_{\RS}^\m$, taking the derivative of the equations of motion we
obtain
\be \pdi\left(\g_\a{\mathcal R}_{\RS}^\a(\hat\psi_\m)-\g_\a
J^\a\right)=0.
\ee
Imposing as initial condition
\be
\left(\g_\a{\mathcal R}_{\RS}^\a(\hat\psi_\m)-\g_\a
J^\a\right)\big|_0=0,
\ee
the last
equation is equivalent to the trace of the RS case.
Thus the propagator that mediates the interaction between two
conserved sources is the same in both cases. In particular we find
\be
\pdi \hat \psi_{\WRS}^\m=J^\m-\frac{1}{2}\g^\m \g_\a
J^\a+\g^\m\xi,
\ee with  $\pdi \xi=0$. The interaction between
 conserved sources can be read from the quantity
\be
\bar J_\m \hat \psi^\m_{\WRS}= \bar
J^\m\frac{1}{\Box}\left(\eta_{\m\n}\pdi+\frac{1}{2}\g_\m\pdi\g_\n\right)J^\n,
\ee
 which coincides with that of the RS (see {\em e.g.}
\cite{Deser:1977ur}). In particular, this guarantees
the unitarity of the theory for conserved sources.

It is interesting to note that, as happens for
the spin-2 Lagrangian, the WRS massive case is different
from the RS one and new degrees of freedom appears in the propagator.

\subsection{Remarks on the Consistent Coupling of the spin-$3/2$ field}

A possible source of problems that appears in the study of theories of
higher spin fields, both massive and massless, is the {\em consistent} coupling of
the field. For the massive spin-$3/2$ field, it was found in
\cite{Johnson:1960vt,Velo:1969bt} that there are problems with
unitarity and causal propagation once the field is coupled to an
external electromagnetic source. For the massless case,
 the inconsistency of the minimal
 coupling of the RS field
to electromagnetism occurs already at an algebraic level. Namely, if
we substitute the ordinary derivative by a covariant derivative in
the RS action, differentiating with the covariant derivative
$D_\m=\pd_\m-\I e A_\m$  and after using
 the Bianchi identity of the free RS action,
$$\pd_\m {\mathcal R}_{\RS}^\m=0,$$
 we find \cite{VanNieuwenhuizen:1981ae}
$$
F_{\m\n}\g^\m\psi^\n=0.
$$
The previous expression
means that either $\psi_\m=0$ or that the photon is a pure gauge
excitation. A similar problem occurs for every massless higher spin
theory, as the Bianchi identities of the free theory always imply
some condition in the background field. It was suggested in
\cite{Skvortsov:2007kz} that the description in terms of
traceless fields may
alleviate this problem as the Bianchi identities are less stringent
in this case.

For the WRS case, coupling minimally the action to
the electromagnetic field, one finds the equations of motion
\be
\label{chargedeom}
\big(\d_\m^\a-\frac{1}{4}\g^\a\g_\m\big)\e^{\m\n\r\s}\g_5\g_\n D_\r \hat
\psi_\s=\I\left(\g^\m D_\m \hat \psi^\a -\frac{1}{2}\g^\a D^\m
\hat\psi_\m\right)=0.
\ee
After applying the covariant derivative, the equations of motion read
\be
eF_{\m\n}\g^\m\hat\psi^\m=\frac{\I}{2}\g^\b D_\b (D_\a \hat\psi^\a),
\ee
which is not a constraint but a field equation\footnote{The same
happens if one considers the coupling of the gauge-fixed RS
action.}. One can show that the previous system of equations is hyperbolic
and that the propagation is causal.

The main concern about the previous
coupling is that the states of low spin corresponding to
$\pd_\a\hat\psi^\a$ are turned on by the interaction, and this may
spoil the unitarity of the theory. The absence of a gauge symmetry implies that the
Slavnov-Taylor identities can not be derived in the standard fashion
and unitarity may be violated even at tree level.
Besides, this implies that no Fadeev-Popov or Nielsen-Kallosh ghosts
will be present in the quantization of the WRS action.
We leave the study of
these issues for future research\footnote{Even if unitarity
 is not preserved, one could try to introduce new fields of spin-$1/2$
to obtain a consistent theory.} \cite{Blas08}.

\section{Supersymmetric Extensions of WTDiff}\label{sectionSugra}

A natural question about the alternative Lagrangian  describing
the propagation of the spin-2 particle (WTDiff) is whether
 it admits a minimal supersymmetric extension.
 Due to the fact that
 the number of \emph{off-shell} and \emph{on-shell}  degrees of freedom of the
massless WTDiff case coincides with that of the standard action for spin-2, one
may wonder about the existence of an action for the spin-$3/2$ field
such that the total {\em minimal} action of the WTDiff graviton plus gravitino has a certain
 global supersymmetry .
A first sign that this may  not be possible is that, as we showed in Section \ref{sectionprop}, the
only Lagrangian for the field $\psi_\m$ that describes purely
spin-$3/2$ \emph{on-shell} is the RS Lagrangian whose
supersymmetric counterpart is the usual linearized Einstein-Hilbert
action\footnote{We could consider actions for the bosonic sector
with more degrees of freedom \emph{e.g.} allowing for a propagating
torsion or non-metricity, but this goes beyond the present work.}.
One may still think that the supersymmetric transformations
can be deformed so that RS action admits another supersymmetric partner.
We will study this possibility in a completely
generic way.\\

Let us first write the form of the WTDiff action
in four dimensions \cite{Alvarez:2006uu},
\be
 {\mathcal{S}}_{{\WTDiff}}=\int \di^4x \left(\frac{1}{4}\partial_\mu
       \hat h^{\nu\rho}\partial^\mu \hat h_{\nu\rho}-\frac{1}{2}
       \partial_\mu \hat h^{\mu\rho}\partial_\nu h^\nu_\rho\right).
\ee
Under a general variation $\d h_{\m\n}$ the
action changes as,
\ba
&&\d{\mathcal S}_{\WTDiff}^{(2)}=\int \di^4 x\ \delta \hat
h_{\m\n}\left(R_{\m\n}^L(\hat h) -\frac{1}{2}\eta_{\m\n} R^L(\hat
h)\right)=
\nonumber\\
&&\frac{1}{4}\int \di^4 x\ \delta h_{\m\n} \left(4\eta^{\a
b}\eta^{\b(\m}\eta^{\n)a} -2\eta^{\a\b}\eta^{a\m}\eta^{b\n}-
\eta^{ab}\eta^{\m\a} \eta^{\n\b} - \eta^{\m\n}\left\{\eta^{\a
a}\eta^{\b b}-\frac{3}{4}\eta^{a b}\eta^{\a\b}\right\} \right)\pd_\a
\pd_\b h_{ab}.\nonumber \ea For the spin-$3/2$  Majorana field
$\psi_\m$ we will take the general action (\ref{gaction}).
 The most general supersymmetric transformation for Majorana spinors and gravitons
 can be written as\footnote{The supersymmetric transformation should preserve
the traceless condition, which for the usual supersymmetric transformation of the
graviton with $A=0$ in (\ref{generaltr}),
implies
\be \d h=\bar \e \g^\m \psi_\m=0.
\ee
This seems to
imply that the supersymmetric partner of the field $\hat h_{\m\n}$
should be
 the field $\hat \psi_\m$ but, as we will see, this is not the case.}
\ba
\label{generaltr}
\delta h_{\m\n}&=&\bar\epsilon \g_{(\m}\psi_{\n)}+A \eta_{\m\n}\bar\epsilon\g^\r\psi_\r,\nonumber\\
\delta \psi_\m&=&\left(B\pd_\m h+C\pd_a h^a_\m+D\g_\m \g^\n \pd_\n h
+E \g_\m \g^\a\pd_b h^b_\a+ F\s^{ab}\pd_a h_{\m b}\right)\e. \ea
Some of the previous transformations are simply field redefinitions
or gauge transformations
for certain Lagrangians but we will just consider all the coefficients as independent.\\

The variation of the bosonic Lagrangian can be written as \ba
\label{susygr} \delta{\mathcal S}_{\WTDiff}^{(2)}=\frac{1}{4} \int
\di^4 x\ \bar \e\Big(-\eta^{ab}\g^\a\psi^\b&&+2\eta^{\a
a}\g^b\psi^\b+2\eta^{\a a}\g^\b\psi^b
-2\eta^{\a\b}\g^b\psi^a\nonumber\\
&&+\frac{3}{4}\eta^{ab}\eta^{\a\b}\g^\r\psi_\r -\eta^{\a a}\eta^{\b
b}\g^\r\psi_\r\Big)\pd_\a \pd_\b h_{ab}. \ea For the variation of
the fermionic part  we find \ba \label{susyrs}
&&\d {\mathcal S}^{(3/2)}=\nonumber\\
&&-\int\di^4 x \ \bar \e
\Big((2B(\l+\zeta)+4D(2\l+\zeta)-F\l)\eta^{ab}\g^\a\psi^\b
+(2C\l+4E(2\l+\zeta)+F\l)\eta^{\a a}\g^b\psi^\b\nonumber\\
&&+\zeta(2C-F)\eta^{\a a}\g^\b\psi^b +F\zeta\eta^{\a\b}\g^b\psi^a
+(2B(\l+\vartheta)+2D(\l+4\vartheta-\zeta)-F\vartheta)\eta^{ab}\eta^{\b\a}\g^\r\psi_\r
\nonumber\\
&&+(2C\vartheta+2E(\l+4\vartheta-\zeta)+F(\l+\vartheta))\eta^{\a
a}\g^b\g^\b\g^\r\psi_\r +\l(2C-F)\eta^{\a a}\eta^{\b b}\g^\r\psi_\r
\Big)\pd_\a \pd_\b h_{ab}. \ea Comparing the third and forth
coefficients of (\ref{susygr}) and (\ref{susyrs}), we find $C=0$.
From the relation between the last coefficient  and the forth one of
(\ref{susygr}), we find $\zeta=-2\l$. Finally, comparing the second
and forth coefficient we arrive at $F\zeta=0$, and thus there is no
way in which both variations can cancel each other. Thus, we
conclude that there is no minimal supersymmetric system
including the WTDiff Lagrangian.

One could try to add some constraints to the action as
was done in \cite{Nishino:2001gd}, to find a supersymmetric
action. However, the addition of these Lagrange
multipliers goes beyond the
minimal coupling and
can be problematic \cite{Gabadadze:2005uq}.

%
\section{Conclusions}

In this work, we have studied the possibility of describing the
free spin-$3/2$ field by a Lagrangian different than the usual Rarita-Schwinger
Lagrangian \cite{Rarita:1941mf}. From the fact that the covariant
fields have more components than the two physical components
of the field $\pm 3/2$, we expect that the action that describes
consistent spin-$3/2$ will be endowed with a gauge symmetry to
make the extra polarizations spurious.

For the rest of the cases,
we expect the theory to be {\em non-unitary} but we did not study
the general case in detail. Instead, we identified the possible gauge symmetries
and found the actions which are invariant under these transformations. It turns
out that there are just two possibilities: the RS action (and all the actions
related to it by a field redefinition) and a new action endowed with a $S$-symmetry.
We have called this alternative possibility WRS, as the $S$-symmetry is similar to
a Weyl transformation for the spin-2 field (cf. (\ref{generalgauge3/2})) and it
is related to the RS action.

For the WRS action, the study of the equations of motion reveals that,
apart from the spin-$3/2$ polarizations, it also includes spin-$1/2$ degrees of freedom.
Nevertheless, and as happens whenever one fixes the gauge through a covariant gauge
fixing term, once coupled to conserved sources
and endowed with the appropriate initial conditions, both the
WRS and the RS theories give the same physical predictions. Indeed,
we showed that the propagator which
describes the interaction between two conserved sources coincides in both cases.
In other words, both theories coincide once the extra degree of freedom is
integrated out.

The introduction of interaction is always a delicate point for higher spin fields.
The difficulties come from the fact that the presence of interaction typically
destroys the gauge invariance of the theory. This may imply certain
conditions on the background field which may render it trivial or the propagation of
low spin modes which may spoil unitarity. For the usual massless spin-$3/2$ field,
when one tries to couple it to electromagnetism, the presence of
a non trivial field $\psi_\m$ means that the electromagnetic background must be trivial.
 This problem is not present in the WRS
 case, but, as expected, the interaction turns on the low spin modes.
 We leave the study of the unitarity of this coupling for future research
 \cite{Blas08}.\\

Finally, we studied the possibility of finding a supersymmetric action built
out of the WTDiff Lagrangian for spin-2. We found that the minimal possibility
consisting on the addition of a certain action for a spin-$3/2$ field does not
work. This means that the linear WTDiff action has not a (minimal)
supersymmetric extension. This result is interesting because  supersymmetry
plays a key role in finding a consistent coupling of the spin-$3/2$ field
\cite{Deser:1976eh,VanNieuwenhuizen:1981ae}.
The previous result naively implies that gravity described from the WTDiff action
can not be coupled to a spin-$3/2$. However, recall that the problem
with the consistent coupling of the spin-$3/2$ is that the action should
include a gauge invariance coming from a
deformation of the gauge invariance of the linear action. This is the case for the
coupling of the RS action to gravity where the Bianchi identities associated to the
supersymmetric transformation are automatically satisfied once the Einstein's equations
are imposed \cite{VanNieuwenhuizen:1981ae}.

This seems to indicate that if one couples the spin-$3/2$ field in the same fashion to the
WTDiff vielbein\footnote{It is important to notice that the first order
formulation is also valid for the WTDiff Lagrangian of \cite{Alvarez:2006uu} (see also
\cite{Blas:2007pp}).},
$$
\hat e_{\phantom{a}\m}^{a}=e^{-1/4}e_{\phantom{a}\m}^{a},
$$
one will get a consistent theory once {\em all} the Einstein's equations are
satisfied, which would mean that the integration constant that remains free in the WTDiff
theory would be fixed to zero. This means that even in the absence of supersymmetry,
the cosmological constant would cancel in this case. Whether this naive
expectation holds or not is currently under research \cite{Blas08}.

\section{Acknowledgements}

It is a pleasure to thank D. Diego for many discussions and initial collaboration
on these issues. I would also like to thank J. Russo and the participants of Peyresq XII for
interesting comments.  I am specially grateful to D. Arteaga, G. P\'erez-Nadal, A. Roura and
G. Gabadadze for clarifying discussions. This
work has been partially supported by CICYT grants
FPA-2004-04582, FPA2007-66665C02-02
and
DURSI 2005-SGR-00082.


\end{document}